\documentclass[a4paper,10pt,twoside]{cpc-hepnp}
\usepackage{CJK,upgreek,fancyhdr}
\usepackage{multicol}
\usepackage{graphicx}
\usepackage{subfigure}
\usepackage{booktabs}
\usepackage{amssymb,bm,mathrsfs,bbm,amscd}
\usepackage[tbtags]{amsmath}
\usepackage{lastpage}
\usepackage{pdfpages}

\begin{document}
\begin{CJK*}{GB}{gbsn}
%\begin{CJK*}{GBK}{song}

\fancyhead[c]{\small Chinese Physics C   Vol. 41, No. 1 (2017) 016001}
\fancyfoot[C]{\small 010201-\thepage}

\footnotetext[0]{Received 31 June 2015}

\title{Temperature dependence of the plastic scintillator detector for DAMPE\thanks{Supported by the Strategic Priority Research Program on Space Science of the Chinese Academy of Sciences (Grant No.~XDA04040202-3) and the Youth Innovation Promotion Association, CAS}}

\author{%
      Zhao-Min Wang$^{1,2;1)}$\email{zmwang@impcas.ac.cn}%
\quad Yu-Hong Yu$^{1;2)}$\email{yuyuhong@impcas.ac.cn}%
\quad Zhi-Yu Sun$^{1}$
\quad Ke Yue$^{1}$
\quad Duo Yan$^{1}$\\
\quad Yong-Jie Zhang$^{1}$
\quad Yong Zhou$^{1,2}$
\quad Fang Fang$^{1,3}$
\quad Wen-Xue Huang$^{1}$\\
\quad Jun-Ling Chen$^{1}$
}
\maketitle

\address{%
$^1$ (Institute of Modern Physics,Chinese Academy of Sciences,Lanzhou 730000,China)\\
$^2$ (University of Chinese Academy of Sciences, Beijing 100049, China)\\
$^3$ (University of Science and Technology of China, Hefei 230026, China)\\
}

\begin{abstract}
The Plastic Scintillator Detector (PSD) is one of the main sub-detectors in the DArk Matter Particle Explorer (DAMPE) project. It will be operated over a large temperature range from -$10$ to $30^{\circ}$C, so the temperature effect of the whole detection system should be studied in detail. The temperature dependence of the PSD system is mainly contributed by the three parts: the plastic scintillator bar, the photomultiplier tube (PMT), and the Front End Electronics (FEE). These three parts have been studied in detail and the contribution of each part has been obtained and discussed. The temperature coefficient of the PMT is $-0.320(\pm0.033)\%/^{\circ}$C, and the coefficient of the plastic scintillator bar is $-0.036(\pm0.038)\%/^{\circ}$C. This result means that after subtracting the FEE pedestal, the variation of the signal amplitude of the PMT-scintillator system due to temperature mainly comes from the PMT, and the plastic scintillator bar is not sensitive to temperature over the operating range. Since the temperature effect cannot be ignored, the temperature dependence of the whole PSD has been also studied and a correction has been made to minimize this effect. The correction result shows that the effect of temperature on the signal amplitude of the PSD system can be suppressed.
\end{abstract}



\footnotetext[0]{\hspace*{-3mm}\raisebox{0.3ex}{$\scriptstyle\copyright$}2013
Chinese Physical Society and the Institute of High Energy Physics
of the Chinese Academy of Sciences and the Institute
of Modern Physics of the Chinese Academy of Sciences and IOP Publishing Ltd}%

\begin{multicols}{2}

\section{Introduction}

With the continuous exploration of the universe, humans have made much progress in knowing its essence. So far, we have learned that the universe is mainly constituted by dark matter (about 27\%) and dark energy (about 68\%). The rest - stars, free hydrogen and helium, heavy elements and neutrinos - count for only about 5\% percent of mass of the universe. Though the existence of dark matter has been accepted, we have limited knowledge of the specific properties that  dark matter particles might have. So exploring for signs of dark matter particles has become a major topic of scientific research.

Space exploration is one of the main ways to search for signs of dark matter. Many scientific satellites have been launched for this goal, such as the Fermi Gamma-ray Space Telescope~\cite{lab1}, PAMELA~\cite{lab2} and AMS-02~\cite{lab3}, and certain progress has been made in these subjects. DAMPE (DArk Matter Particle Explore) is a satellite-borne apparatus to measure electrons and photons with much higher energy resolution, to identify possible dark matter signatures, as well to make new discoveries in high energy gamma astronomy. It has also great potential to study the origin and acceleration of high energy cosmic rays, and how they propagate~\cite{lab4}.

The Plastic Scintillator Detector (PSD) system, the top part of DAMPE, is one of the most important sub-systems to detect incident heavy ions. Its goal is to distinguish not only protons from electrons, but also different incident heavy ions with charge numbers of less than 20.
Figure~\ref{fig-PSD-total} shows the Plastic Scintillator Detector (PSD) system. It consists of 82 plastic scintillator bars, which are arranged in two layers and cover an overall active area of 82 cm $\times$ 82 cm. The scintillator bars, with a dimension of 884 mm $\times$ 28 mm $\times$ 10 mm, are parallel to each other in the same layer and perpendicular to those in the other layer. To avoid the presence of any ineffective detection area, the bars in the same layer are staggered by 0.8 cm. Each bar is coupled  to a photomultiplier (PMT) (Hamamatsu R4443 \cite{lab5}) at each end by a silicon rubber. The analog signals from the readout of the PMTs are first fed to the Front End Electronics (FEE), and then sent to a Data AcQuisition system (DAQ) for data storage and analysis.
Due to the harsh space environment, a thermal design was specialized for this sub-detector system. By using an automatic thermal control system, the temperature for the PSD system can be stabilized in a range between -$10$ and $30 ^{\circ}$C.
\begin{center}
\centering
\includegraphics[width=8cm]{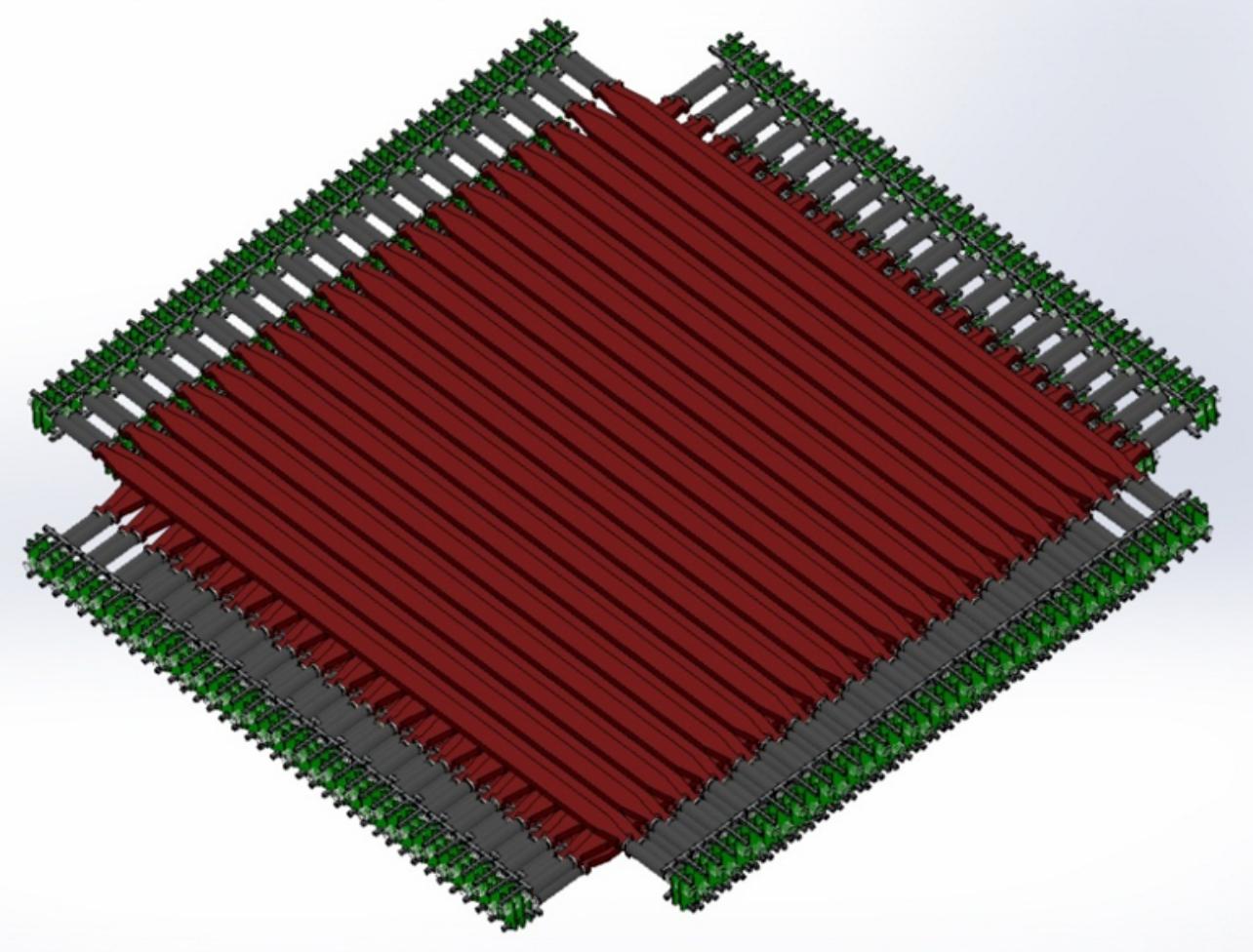}
\figcaption{\label{fig-PSD-total}Plastic scintillator detector (PSD) system }
\end{center}

Because the performance of the PSDs may vary as temperature changes, the temperature dependence has to be studied in detail before the assembly of the whole PSD system. This study also give a method to correct the amplitudes of the output signals at different temperatures.

Most satellite-borne detectors have to face the temperature effect on detector performance, and they mainly focus on studying the temperature dependence of inorganic scintillator combined with the readout PMT~\cite{lab6}\cite{lab7}. Organic plastic scintillator, also read out with PMT, has seldom been studied in recent years.

From analysis, the detection system of the PSD can be divided into three parts: the detection body plastic scintillator bar, the readout PMTs and the FEE. That is, all the temperature effect can be determined by these three elements. Once the assembly of the PSD was finished, the PMTs and plastic scintillator bars would always work together, so it is impossible to study their respective temperature dependence. In this case, an experiment was designed before the assembly began to study these three elements' respective temperature dependence. To figure out each part's contribution to the total temperature coefficient, we chose one PMT, plastic scintillator bar and FEE as the test instrument. In this paper, Section 2 will mainly describe the setup of this experiment. The LED is used to calibrate the PMT, and a $\beta$ source~$^{207}$Bi is used to calibrate the plastic scintillator bar. Such methods have been widely used in studying the temperature dependence of scintillator detectors~\cite{lab6}\cite{lab8}. The test results will be discussed in Section 3. By this experiment, the temperature coefficient of the tested PMT, plastic scintillator bar and FEE could be obtained. According to the results, it is clear that the temperature effect cannot be ignored in data analysis. Since the temperature coefficients of individual PMTs are different from each other, the temperature coefficient obtained above cannot be used to correct the amplitudes of the output signals in different PMTs. So another experiment aimed at studying the whole PSD system with cosmic rays at different temperatures was carried out. This is also a widely used method to calibrate detectors~\cite{lab7}. A temperature coefficient is obtained in this way which can be used in making a temperature correction to minimize the temperature effect, and this will be discussed in Section 4. The temperature coefficients and the way to correct the temperature effect provide the basis for orbit data analysis.

\section{Temperature dependence of the LED, FEE and PMT}

\subsection{Experimental setup}
\label{sec-setup-1}

Figure~\ref{fig-setup-1} shows the experimental setup used for studying temperature dependence of the FEE and the PMT schematically. PMTs and their succesive FEEs were placed into a thermal chamber (Chamber01) and the outputs of the FEEs were sent to a DAQ for data collection.  The PMTs used in the test were the same type and the same batch as those of the PSD system. To match the most probable wavelength of maximum response of the PMTs, a blue LED was used to monitor their gain variations. Since the light intensity from the blue LED is also dependent on the temperature, it is necessary to place it in a room with a constant temperature, thus another thermal chamber (Chamber02) was used to guarantee the stability of light intensity. An optical fiber was used to inject the blue light pulses into the photocathode of the PMTs.  A pulser was used to drive the LED and to trigger the DAQ.
The high voltage (HV) and low voltage direct current (DC) power supplies, the pulser as signal source, the cables, some of the light-guide fiber and the DAQ were placed outside the thermal chambers and kept at room temperature since their performances are not sensitive to temperature.

Both thermal chambers used here were bought from ETOMA Company (type: WSZ62IIITS). They have the same dimensions of 1 m $\times$ 1 m $\times$ 1 m and can be operated in a large dynamic temperature range from $-70^{\circ}$C to $150^{\circ}$C. The  temperature stability is $\pm0.5^{\circ}$C, and the temperature changing speed is faster than $5^{\circ}$C/min.
In order to avoid condensation of water vapour inside the inner chamber, a significant number of desiccants were put into the inner chamber during the measurements. However, the water vapour problem could not be totally eradicated by using desiccant only. When the temperature was below $0^{\circ}$C, some frost could be found on the surface of the tested PMTs. Due to this constraint,the lowest temperature was set at $3^{\circ}$C during the experiment.

The experiment tested the devices in different temperatures to see the changes of their performance. According to previous tests, it needs about 30 minutes for the LED, PMT and plastic scintillator bar to reach equilibrium with the environment temperature. So during the experiment, after the temperature inside Chamber02 reached the set point, we waited for another 60 minutes to ensure thorough thermal equilibrium.
\begin{center}
\centering
\includegraphics[width=8cm]{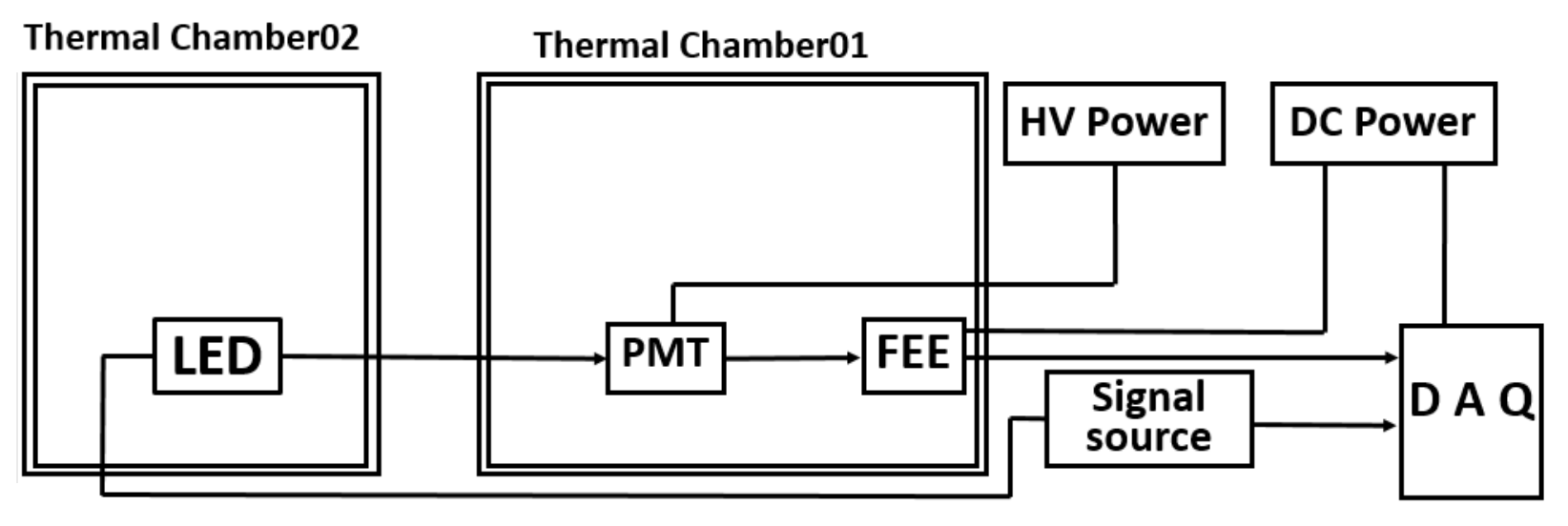}
\figcaption{\label{fig-setup-1}Schematic diagram of the experimental setup for the temperature dependences of the PMT and the FEE. }
\end{center}

\subsection{LED results}

Since the light intensity of the LED varies as the temperature changes, the temperature dependence of the LED was studied first. The temperature inside  Chamber01 in Fig. \ref{fig-setup-1} was kept at $20^{\circ}$C, thus the effect from the FEE and the PMT can be kept the same during the measurements. The temperature inside Chamber02, where the LED was mounted, was changed from $5^{\circ}$C to $35^{\circ}$C with a step of  $5^{\circ}$C.

At a given temperature, the blue LED was turned off and only the rms noise of the electronics was measured first, to check the effect from the FEE. The blue LED was then turned on and the amplitude of the signal was measured. Figure~\ref{fig-led} shows the spectrum recorded at different temperatures of the blue LED after subtracting the effect from the FEE, and the temperature dependence of the LED. By using Gaussian distributions to fit the peaks, the most probable values (MPVs) at different temperatures have been obtained and fitted with a linear function:
\begin{equation}
\label{eq-pled}
P_{LED}=(8.62\pm0.23)T+(195.7\pm5.1),
\end{equation}
where $P_{LED}$ is the ADC channels of the MPV of the pulse signal amplitude for the LED, and $T$ is the temperature of the LED in $^{\circ}$C.

\begin{center}
\centering
\includegraphics[width=6.5cm]{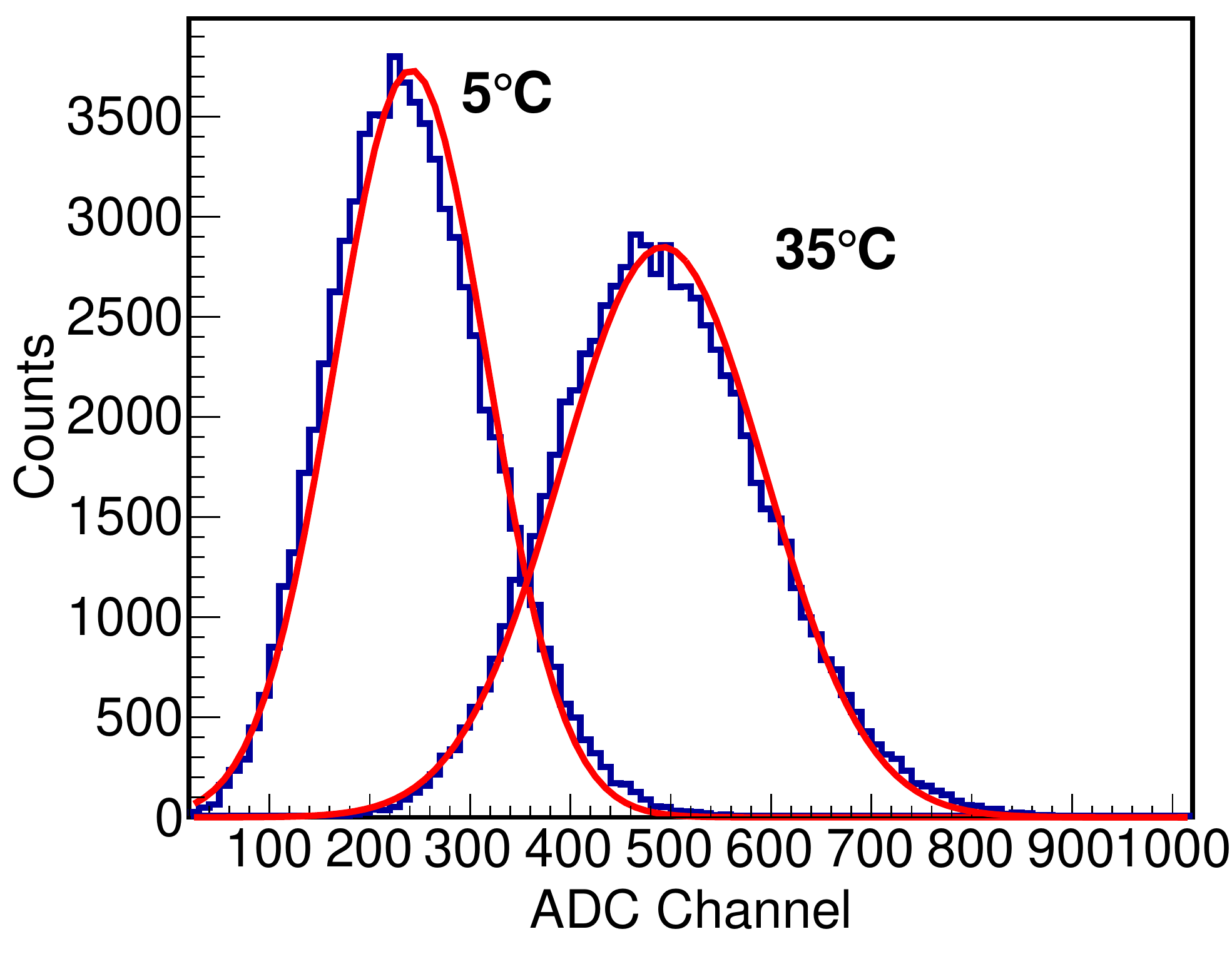}
\includegraphics[width=6.5cm]{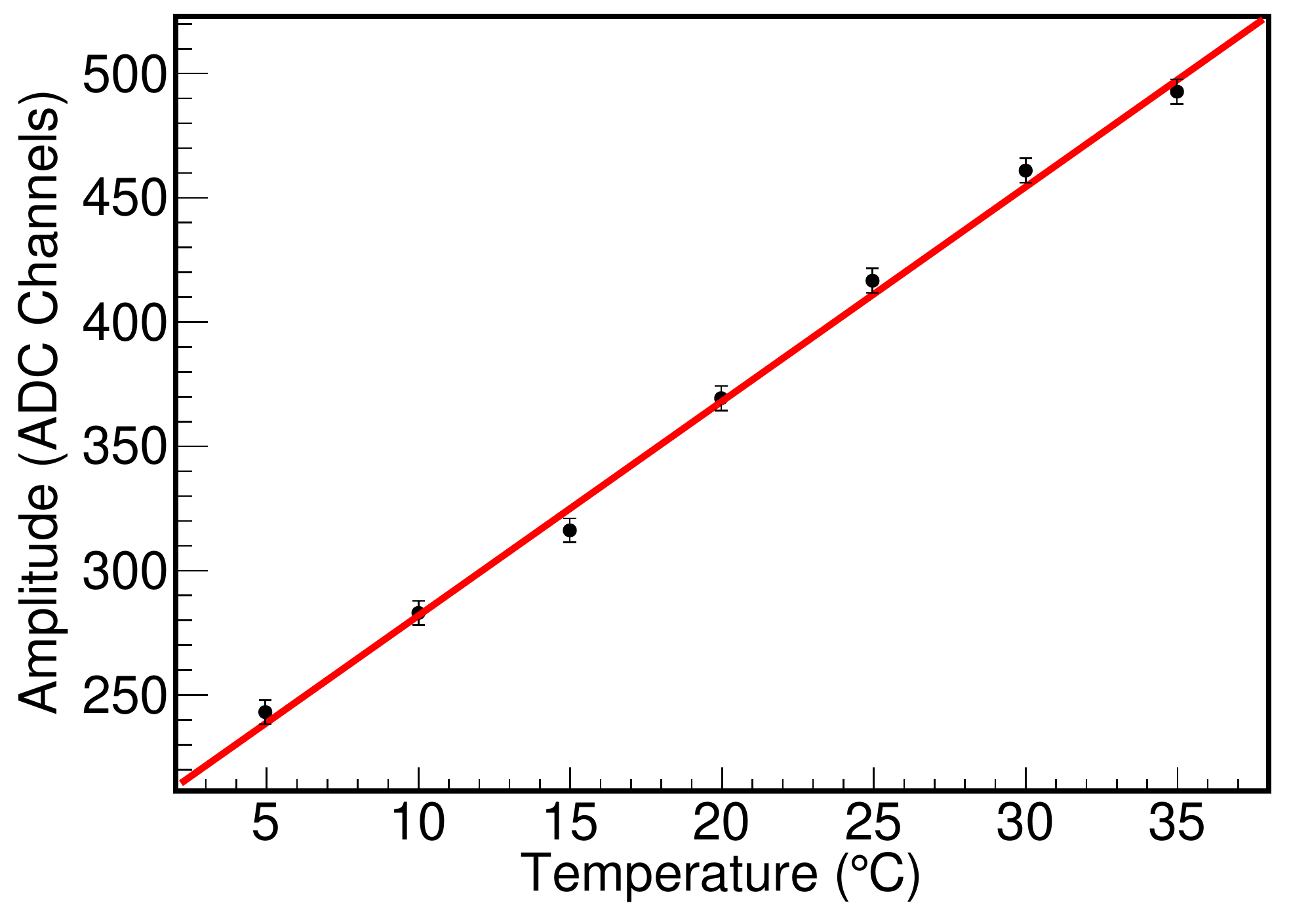}
\figcaption{\label{fig-led} (top) The amplitude spectra of the LED at two different temperatures: $5^{\circ}$C and $35^{\circ}$C. (bottom) The temperature dependence of the LED.}
\end{center}

\subsection{FEE results}

To measure the temperature dependence of the FEE, the temperature inside  Chamber01 in Fig. \ref{fig-setup-1} was changed from $3^{\circ}$C to $42^{\circ}$C with a step of $3^{\circ}$C. At every temperature point, the thermal equilibrium inside the chamber was ensured also. We did not use the output from the blue LED and the PMT as the input of the FEE, but instead we used the output of another pulser, the amplitude of which was kept at a fixed value during all the measurements.

The spectra at different temperatures were recorded and the MPVs were obtained by fitting the spectra with Gaussian distributions. The temperature dependence of the FEE shown in Fig. \ref{fig-fee} can be also fitted with a linear function:
\begin{equation}
\label{eq-pfee}
P_{FEE}=(-0.385\pm0.013)T+(422.7\pm0.3),
\end{equation}
where $P_{FEE}$ is the ADC channels of the MPV of the pulse signal amplitude for the FEE, and $T$ is the temperature of the FEE in $^{\circ}$C. Although it is very small, it cannot be ignored when we think about the effect from varying temperature.

\begin{center}
\centering
\includegraphics[width=7.0cm]{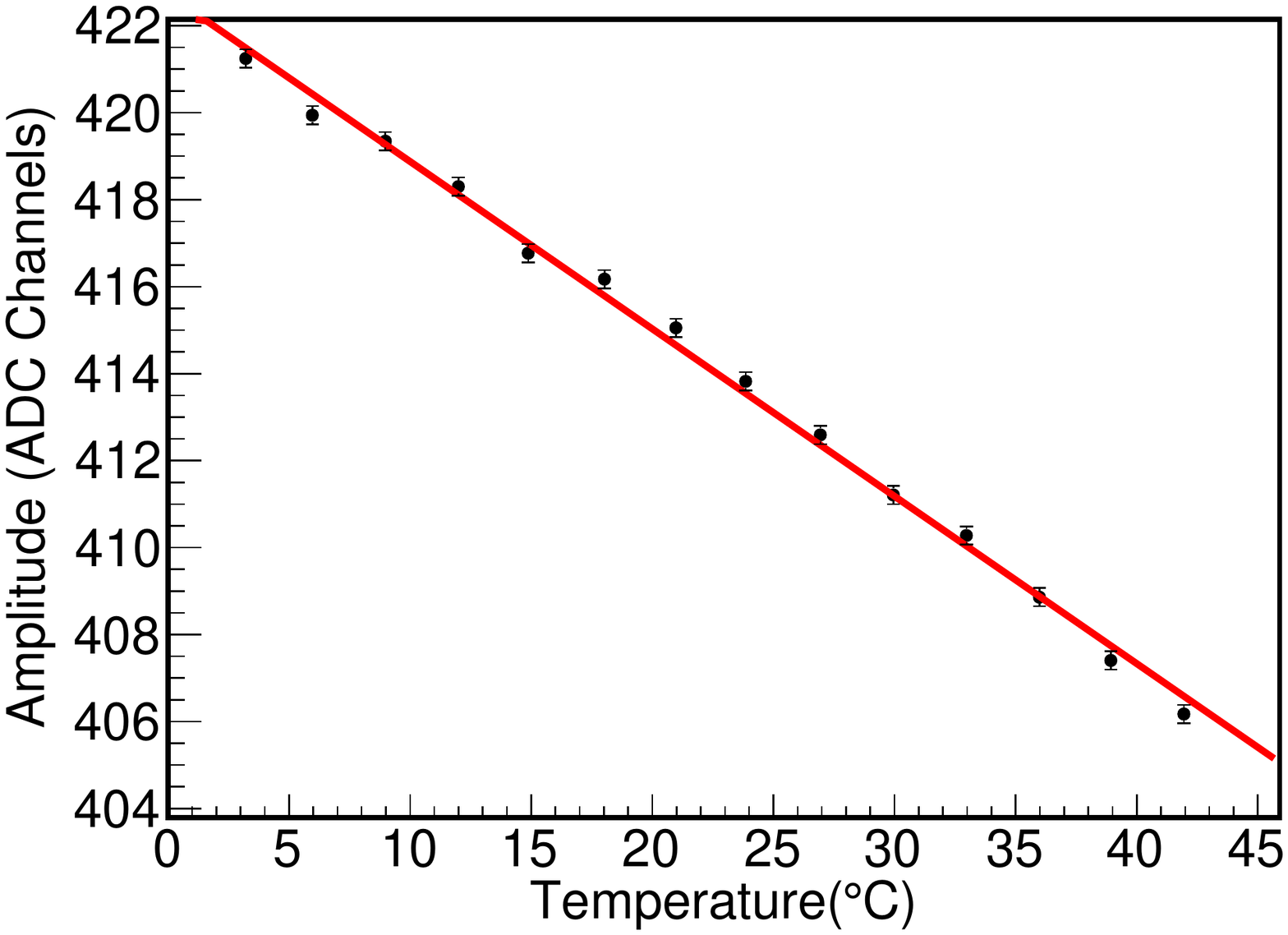}
\figcaption{\label{fig-fee}The temperature dependence of the FEE.}
\end{center}

\subsection{PMT results}
\label{sec-pmt}

During this measurement, the temperature inside Chamber02 in Fig. \ref{fig-setup-1} was set and kept at $20^{\circ}$C to fix the output of the blue LED, while the temperature inside Chamber01 was changed from $3^{\circ}$C to $42^{\circ}$C with a step of $3^{\circ}$C.  At every temperature point, the thermal equilibrium inside the chambers was ensured, as before.

Figure~\ref{fig-pmt} shows the spectrum recorded at different temperatures of the PMT after subtracting the effect from the FEE, and the temperature dependence of the PMT. The MPVs at different temperatures have been obtained and fitted with a linear function:
\begin{equation}
\label{eq-pmpt}
P_{PMT}=(-1.246\pm0.130)T+(414.7\pm3.7),
\end{equation}
where $P_{PMT}$ is the ADC channels of the MPV of the pulse signal amplitude for the PMT, and $T$ is the temperature of the PMT in $^{\circ}$C.

\begin{center}
\centering
\includegraphics[width=7.0cm]{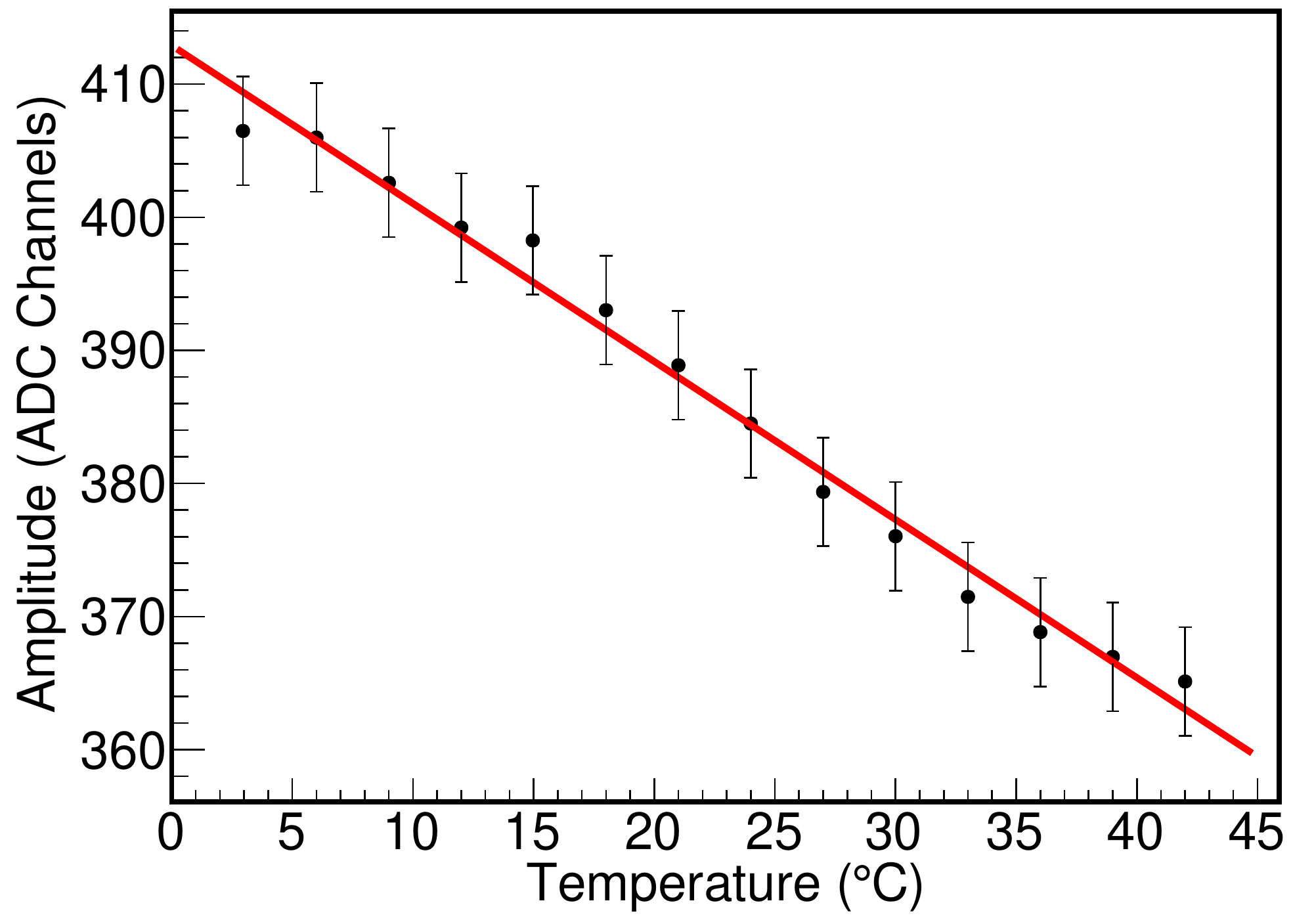}
\figcaption{\label{fig-pmt}The temperature dependence of the tested PMT.}
\end{center}

In Fig.~\ref{fig-pmt}, the errors come from statistical errors of the measured peaks and the temperature instabilities of $\pm0.5^{\circ}$C inside both thermal chambers where the PMT and the LED were settled. Among these, the temperature instability in Chamber02, where the blue LED was mounted, contributed the most to the errors, as can be seen by comparing the slopes in Eq. (\ref{eq-pled})-(\ref{eq-pmpt}).

\section{Temperature dependence of the plastic scintillator bar}

The performance of the plastic scintillator bar is also affected by the variation of the operating temperature~\cite{lab9}. To understand this, a new scheme was set up and $^{207}$Bi was used to replace the blue LED as a source.

\subsection{Experimental setup}
\label{sec-setup-2}

Figure~\ref{fig-setup-2} shows the experimental setup used for studying the temperature dependence of the plastic scintillator bar schematically. A scintillator bar, with dimensions of 180 mm $\times$ 28 mm $\times$ 10 mm, was coupled with a PMT at each end and settled in a thermal chamber (one of the chambers mentioned in Section 2.1). This bar is also from the same batch as those of the PSD system. A $\beta$ source, $^{207}$Bi, was mounted in the middle of the scintillator bar to monitor the amplitude variation, which is induced by the change of the scintillation light output and the relative gain of the PMT. The analog signal from one end of the scintillator bar was fed into a constant discriminator CF8000 which then triggered the DAQ. Meanwhile, the analog signal from the other end of the bar was read out by the same PMT used in the previous measurement in Fig. \ref{fig-setup-1}.

\begin{center}
\centering
\includegraphics[width=8cm]{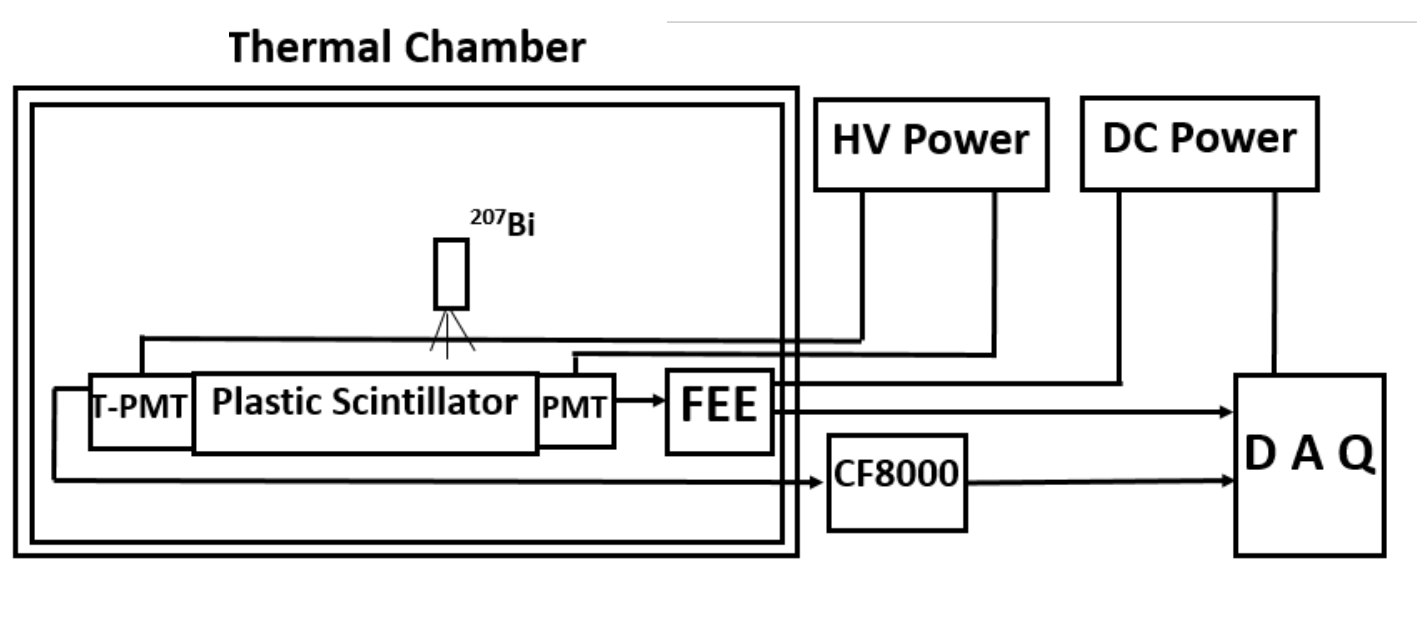}
\figcaption{\label{fig-setup-2}Schematic diagram of the experimental setup for the temperature dependence of the plastic scintillator bar.}
\end{center}

\subsection{Plastic scintillator bar results}

The test procedure is very similar to that mentioned above and was repeated. The temperature inside the chamber in Fig. \ref{fig-setup-2} was changed from $3^{\circ}$C to $42^{\circ}$C with a step of $3^{\circ}$C.

Figure~\ref{fig-psd} shows the spectrum recorded at different temperatures after subtracting the effect from the FEE. The MPVs ($P_{^{207}Bi}$) at different temperatures measured by the $^{207}$Bi source can also be fitted with a linear function:
\begin{equation}
\label{four}
P_{^{207}Bi}=(-4.728\pm0.067)T+(1423.2\pm3.7).
\end{equation}
$P_{^{207}Bi}$ here denotes the ADC channels of the MPV of the $^{207}$Bi source, and $T$ is the temperature of the PMT and plastic scintillator bar in $^{\circ}$C.
\begin{center}
\centering
\includegraphics[width=7.0cm]{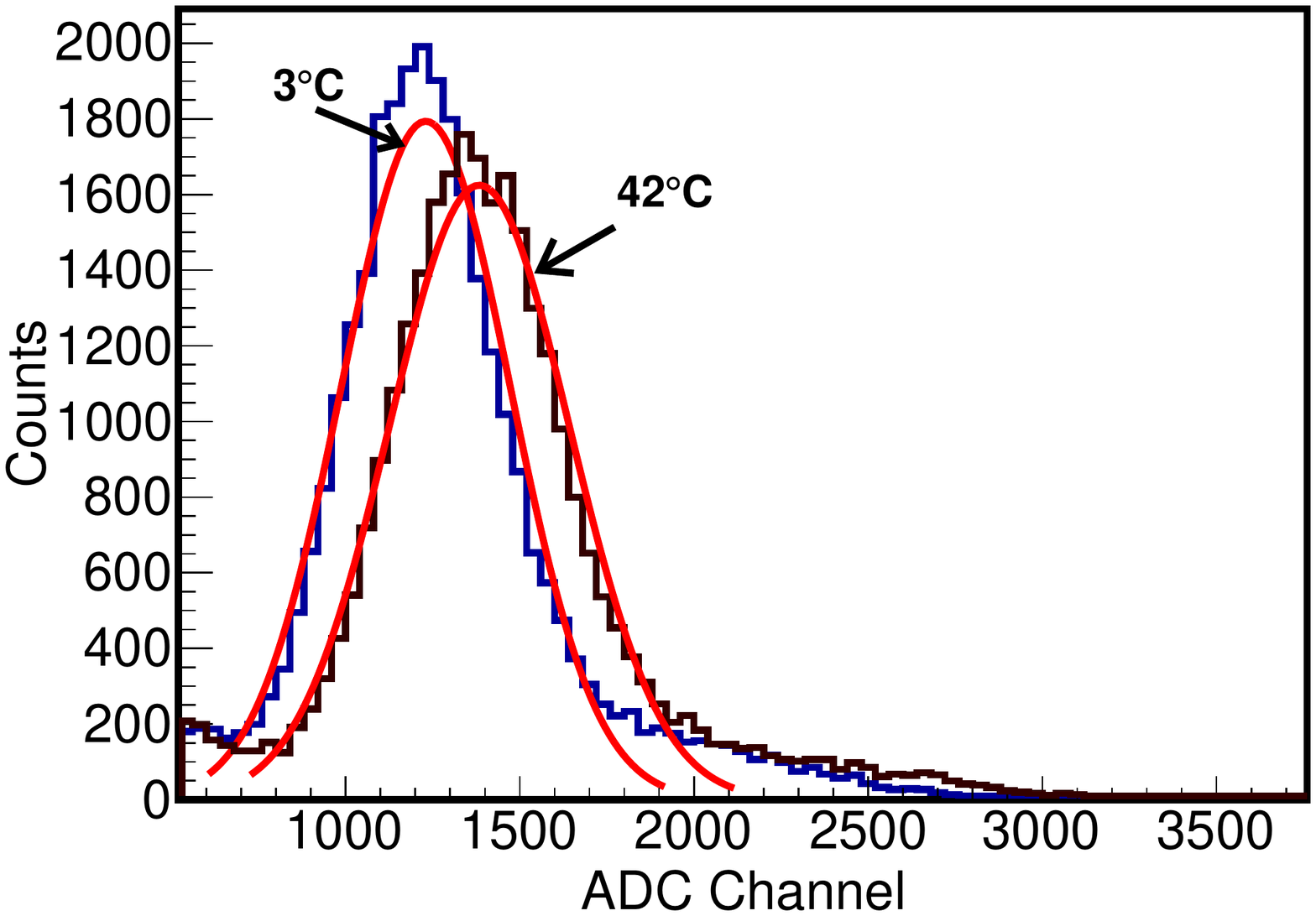}
\figcaption{\label{fig-psd}The amplitude spectra of the $\beta$ source $^{207}$Bi at two different temperatures, $3^{\circ}$C and $42^{\circ}$C.}
\end{center}

To obtain the temperature dependence of the scintillator bar, the effect from the PMT must be subtracted. Since the same PMT and the same FEE were used and their temperature dependence had been studied already, the temperature dependence of the scintillator bar can be deduced as follows.

The temperature coefficient \cite{lab9} can be defined as:
\begin{equation}
\label{five}
C = \frac{S}{P(T=20^{\circ}C)},
\end{equation}
where $S$ is the slope of the MPV curve, and $P(T=20^{\circ}C)$ is the ADC channels of the MPV at a temperature of $20^{\circ}$C (room temperature). Thus, those of the PMT and the combination of the PMT and the scintillator bar are calculated by using Eq. (3) and (4):
\begin{equation}
\label{six}
C_{{\rm {PMT}}} = (-0.320\pm0.033)\%/^{\circ}C,
\end{equation}
\begin{equation}
\label{seven}
C_{{\rm {scintillator}+\rm {PMT}}} = (-0.356\pm0.005)\%/^{\circ}C.
\end{equation}
Ignoring the tiny interaction between the PMT and the scintillator bar, the coefficient of the plastic scintillator bar is obtained:
\begin{align}
\label{eight}
C_{{\rm {scintillator}}}&=C_{{\rm {scintillator}+\rm {PMT}}}-C_{{\rm {PMT}}} \nonumber \\
&= (-0.036\pm0.038)\%/^{\circ}C.
\end{align}

The temperature coefficient of the plastic scintillator bar is much smaller than that of the PMT in the test temperature range. This indicates that the variation of signal amplitude mainly comes from the PMT, and the scintillation light output of the organic plastic scintillator is not very sensitive in the range of normal operation temperature.

\section{Correction of the PSD system}

From the previous measurement and analysis, the temperature dependence of the plastic scintillator bar is tiny compared with that of the PMT. Since the scintillator bar used in the previous test is from the same batch of  scintillator bars assembled for the PSD system, according to Eq. (8) it can be deduced that the temperature coefficient is around $-0.036\%/^{\circ}C$. The difficulty for the whole PSD system comes from the readout PMTs: there are 164 readout PMTs in total, and each of the PMTs is different. The differences make the temperature coefficient of the PMT obtained in the previous test useless. It is essential to gather the information of the temperature coefficient of each readout channel. Besides, the experiments described in Sections 2 and 3 only tested the devices in the temperature range from $3$ to $42^{\circ}$C. In order to cover the temperature range from -$10$ to $30^{\circ}$C, a more advanced thermal chamber was used which could use nitrogen gas as the inner circulation gas to expel the water vapour to a great extent. With this chamber, a similar test to the single plastic scintillator bar but using cosmic rays was implemented for the whole PSD system with the temperature ranging from -$10$ to $30^{\circ}$C. All the values of the temperature coefficients were obtained and kept in a housekeeping database. These coefficients will be used in the following correction work.

To grasp the temperature dependence performance of the whole PSD system before space launch, a continuously changing temperature condition was adopted to simulate the harsh environment in orbit, and find a way of correction to minimize this temperature effect. The whole PSD system was no longer operated at a fixed temperature. It was placed inside a variable thermal-vacuum machine KM5. The temperature inside the thermal-vacuum machine varied, but slowly. The temperature varied from $20^{\circ}$C to about -$2^{\circ}$C, over about 1500 minutes. To monitor the change of the whole detection system, several thermistors were embedded into the PSD during fabrication. The test was also performed with cosmic rays. The scientific data from cosmic rays were acquired in event-by-event mode, and a time tag was also recorded at the same time as a good event happening. The temperature values from different monitor thermistors were recorded at a rate of $1 count/16 s$. It is not difficult to align these two time tags. The temperature variation line and the profile graph of the cosmic ray signal amplitude at the same time is shown in Fig. \ref{psd-variation}. A linear function is used to describe the trend in this graph. According to the slope of the fitting function, the amplitude of the signals increases by about 2.1 ADC channels per hour, which was caused by the cooling temperature.
\begin{center}
\centering
\includegraphics[width=7cm]{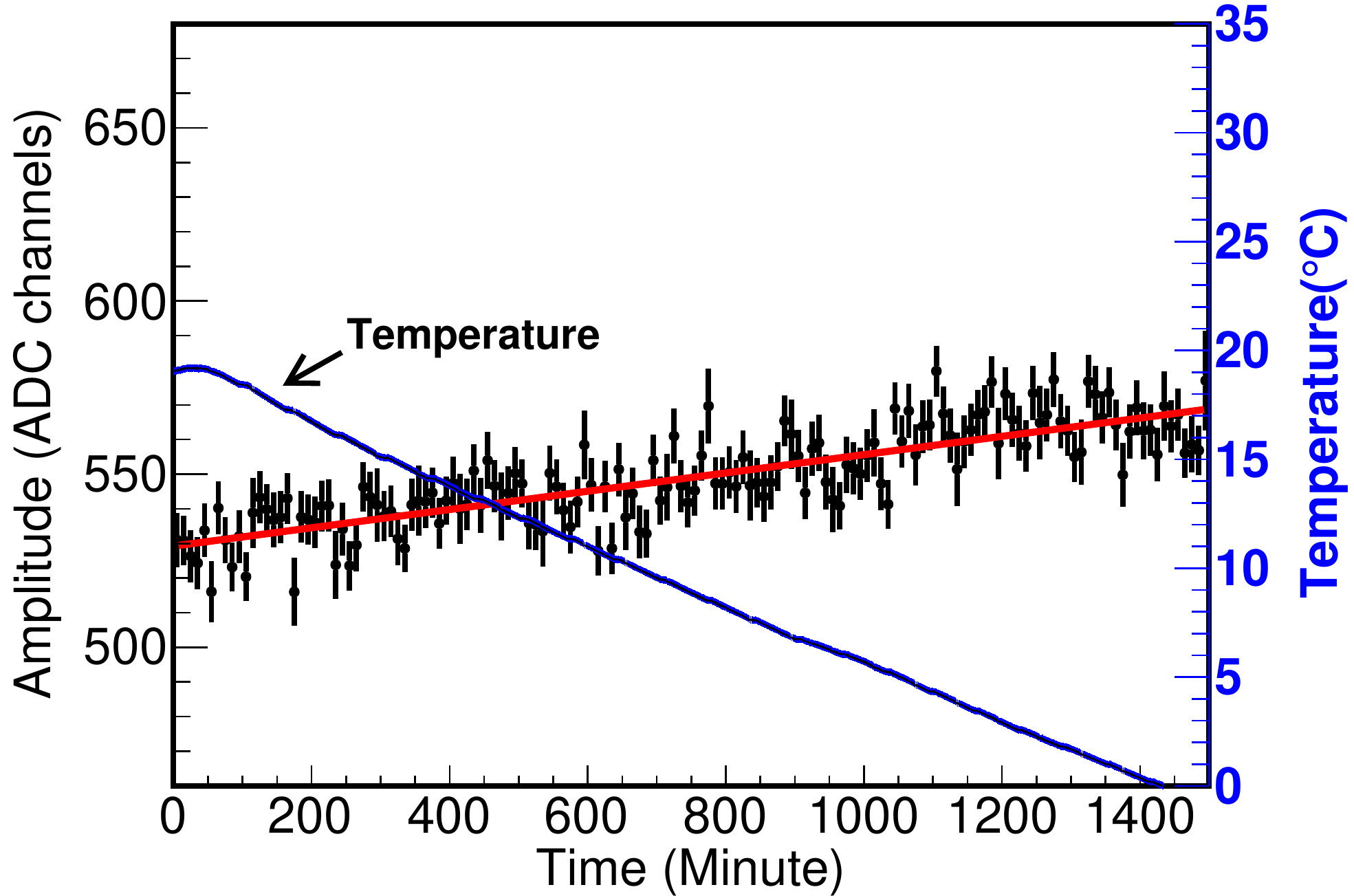}
\figcaption{\label{psd-variation} Temperature variation and profile graph of the cosmic ray signal amplitudes.}
\end{center}

In order to minimize the influence of the variable temperature, a correction is performed as follows.

For each incident particle event, the correction function below is used to correct the data back to the situation for $20^{\circ}$C.
\begin{equation}
\label{modification1}
A_{{\rm {after}}} =A_{{\rm {before}}}-Co*(T-20),
\end{equation}
where $A_{before}$ is the ADC channels of each event that has not been corrected yet, and $A_{after}$ is that for the corrected event. $Co$ is the collected coefficient data from the before test with a value of -2.68, and $T$($^{\circ}$C) is the temperature of the monitor thermistors related to the readout channels. The FEE pedestal has already been subtracted from $A_{before}$. From the previous test, we know that the pedestal is also temperature dependent, so considering the pedestal, the temperature correction can be presented as:
\begin{equation}
\label{modification2}
A_{{\rm {before}}} =A_{{\rm {original}}}-(Ped_{T=20}-Ped_{co}*(T-20)),
\end{equation}
where $A_{original}$ is the ADC channels of each event from which the pedestal  has not been subtracted, $Ped_{T=20}$ is the MPV of the pedestal at $20^{\circ}$C, and $Ped_{co}$ is the slope of pedestal signal variation, with a value of -0.75.

After correction, the profile graph of cosmic ray signal amplitudes is shown in Fig. \ref{psd-modification}. Compared with the profile graph in Fig.~\ref{psd-variation}, the amplitude of the signal variation is minimal, and the slope of the fitting function decreases from $0.0347(\pm0.0015)$ to $0.0001(\pm0.0015)$. This means that the amplitude of the signals will remain stable after correction, and the temperature coefficient values obtained from the whole PSD system will be useful for future correction in the orbit.
\begin{center}
\centering
\includegraphics[width=6.8cm]{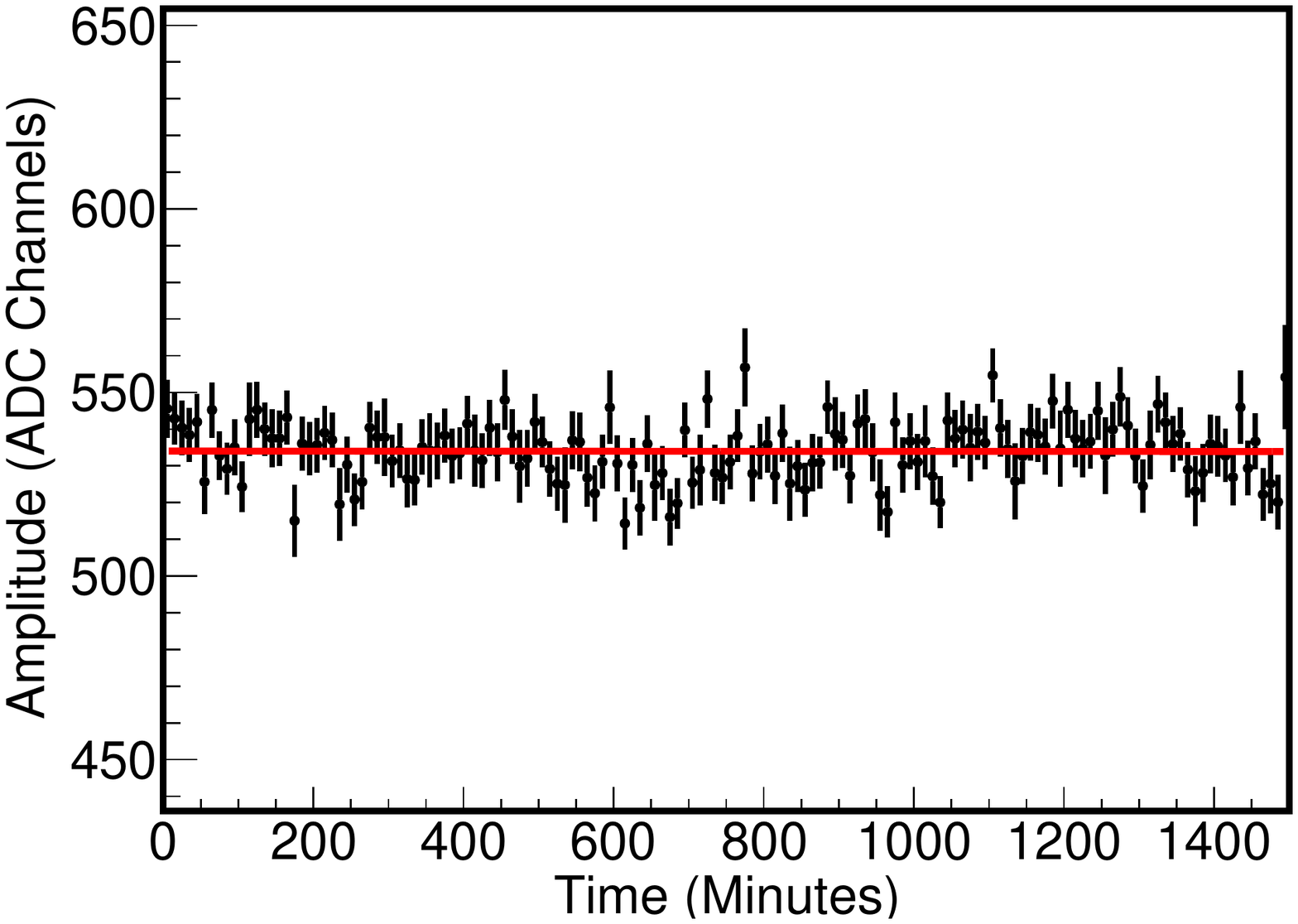}
\figcaption{\label{psd-modification}Profile graph of the cosmic ray signal amplitudes after correction.}
\end{center}

\section{Summary}

This paper is mainly focused on the temperature dependence and correction of the PSD system. After analysis, the plastic scintillator bar, readout PMTs and FEE are the main elements, and measurements were performed in the range $12^{\circ}$C to $42^{\circ}$C to avoid water vapor condensation. The temperature dependence of the blue LED which is used to test the PMT was also studied in an advanced test. The temperature coefficient of the PMT from the same batch of the whole PSD system is found to be $-0.320(\pm0.033)\%/^{\circ}$C, while the coefficient of plastic scintillator bar is only $-0.036(\pm0.038)\%/^{\circ}$C. This result shows that the readout PMT is the most important factor to contribute to the temperature effect. To grasp the temperature dependence performance of the detection system before space launch, all of the temperature coefficient values of the whole PSD have been gathered, and a continuously changing temperature condition was used to simulate the harsh environment in orbit. A correction has been found and used to eliminate this temperature effect by using the gathered coefficient values. After the correction, the signal amplitudes remain stable as the temperature changes. In other words, with this correction, the effect of temperature on the signal amplitude of the PSD system can be suppressed.

\end{multicols}
\vspace{-1mm}
\centerline{\rule{80mm}{0.1pt}}
\vspace{2mm}

\begin{multicols}{2}

\end{multicols}

\clearpage
\end{CJK*}

\begin{thebibliography}{90}

\vspace{3mm}
\bibitem{lab1}A.A.Abdo, M.Ackermann, M.Ajello, et al, Phys.Rev.Lett, \textbf{102}(18):181101(2009)
\bibitem{lab2}O.Adriani, G.C.Barbarino, G.A.Bazilevskaya, et al, Nature, \textbf{458}(7238):607-609(2009)
\bibitem{lab3}M.Aguilar, G.Alberti, B.Alpat, et al, Phys.Rev.Lett, \textbf{110}(14): 1948-1954(2013)
\bibitem{lab4}J.Chang, Journal of Engineering Studies, \textbf{2}(2):95-99(2010) (in Chinese)
\bibitem{lab5}http://www.hamamatsu.com/resources/pdf/etd/PMT\_TPMZ\\0002E.pdf, retrieved 5th March 2016
\bibitem{lab6}P.L.Wang, Y.L.Zhang, Z.Z.Xu \& X.L.Wang, SCIENCE CHINA, Physics, Mechanics\&Astronomy,  \textbf{57}(10): 1898-1901(2014)
\bibitem{lab7}J.Isbert, Jr.J.H.Adams, H.S.Ahn, et al, Advances in Space Research, \textbf{42}(3): 437-441(2008)
\bibitem{lab8}G.Barbarino, M.Boschcrini, D.Campana, et al, Nuclear Physics B, \textbf{125}(3): 298-302(2003)
\bibitem{lab9}J.F.CAMERON, C.G.Clayton, R.A.Spackman et al, Nuclear Electronics, \textbf{1}:95-110(1962)


\end{thebibliography}
\end{document}